\documentclass{aastex701}

\usepackage{hyperref}

\begin{document}

\title{Lightstack: A Python Package for Creating Photometric Data Cubes}

\author[orcid=0000-0001-6687-0402,sname='Wille']{Andressa Wille}
\affiliation{Instituto de Física, Universidade Federal do Rio Grande do Sul, Porto Alegre, RS 90040-060, Brazil}
\email[show]{andressa.wille@ufrgs.br} 

\author[orcid=0000-0001-7207-4584, gname='Rafael',sname='S. de Souza']{Rafael S. de Souza}
\affiliation{Centre for Astrophysics Research, University of Hertfordshire, College Lane, Hatfield, AL10 9AB, UK}
\affiliation{Instituto de Física, Universidade Federal do Rio Grande do Sul, Porto Alegre, RS 90040-060, Brazil}
\affiliation{Department of Physics \& Astronomy, University of North Carolina at Chapel Hill, NC 27599-3255, USA}
\email{rd23aag@herts.ac.uk}

\author[orcid=0000-0003-3220-0165,gname='Ana L.',sname='Chies-Santos']{Ana L. Chies-Santos}
\affiliation{Instituto de Física, Universidade Federal do Rio Grande do Sul, Porto Alegre, RS 90040-060, Brazil}
\email{ana.chies@ufrgs.br}

\author[orcid=0000-0001-6540-0767,gname='Thallis',sname='Pessi']{Thallis Pessi}
\affiliation{European Southern Observatory, Alonso de Córdova 3107, Vitacura, Casilla 19001, Santiago, Chile}
\email{thallis.pessi@gmail.com}

\author[orcid=0000-0002-0406-076X,gname='Emille',sname='E. O. Ishida']{Emille E. O. Ishida} 
\affiliation{Universit\'e Clermont Auvergne, CNRS/IN2P3, LPC, F-63000 Clermont-Ferrand, France}
\email{emille.ishida@CLERMONT.IN2P3.FR}

\author[orcid=0000-0002-2308-6623,gname='Alberto',sname='Krone-Martin']{Alberto Krone-Martins} 
\affiliation{Donald Bren School of Information and Computer Sciences, University of California, Irvine, CA 92697, USA}
\email{algolkm@gmail.com}

\begin{abstract}

Multi-band photometry traces diverse physical processes across a wide range of wavelengths. In recent decades, this field has been driven by the rapid growth of multi-imaging datasets, from high-resolution observation from Hubble Space Telescope and James Webb Space Telescope to the forthcoming large-scale surveys enabled by the Roman Space Telescope and Rubin Observatory, for example.
In this work, we present \textsc{lightstack}, a Python package for combining standalone images into photometric data cubes. The workflow consists of three main steps: cropping a region of interest from a mosaic across all available filters; stacking the images to construct the data cube; and performing PSF matching on the cube. 
This package is intended for preparing data for studies involving multi-band photometry. The code is released under an MIT license and is available on \href{https://github.com/AndressaWille/lightstack}{GitHub} together with a Jupyter tutorial notebook. The version used for this
publication (v0.2.1) is archived on
\href{https://doi.org/10.5281/zenodo.20360028}{Zenodo}.

\end{abstract}


\section{Introduction} 

Multi-band photometric imaging has become central to modern astrophysics, providing wavelength-dependent information that forms the basis of spectral energy distribution (SED) fitting \citep[e.g.,][]{maraston2005, 2009Conroy, vazdekis2016}, which is used to infer physical properties such as stellar mass, age, metallicity, dust attenuation, and star formation rate. Over the past decades, this field has been transformed by the rapid growth of deep, high-resolution multi-band imaging datasets, first from the Hubble Space Telescope (HST) and, more recently, from the James Webb Space Telescope (JWST). Both facilities provide observations in dozens of filters spanning the ultraviolet to infrared regimes.  

In the coming years, with a wider field of view, though fewer filters, Nancy Grace Roman Space Telescope \citep{akeson2019} will also produce valuable multi-band data. Ground-based surveys, such as the Javalambre Physics of the Accelerating Universe Astrophysical Survey \citep[J-PAS;][]{bonoli2021} will provide high-precision photometry with 54 narrow-band filters, and the Legacy Survey of Space and Time \citep[LSST;][]{ivezic2019} at the Vera Rubin Observatory, will map the sky and transform time-domain astronomy, producing a massive amount of data in six optical filters. With such rich datasets available, it is essential to have tools for handling the high spatial resolution combined with the low-resolution SED that multi-band photometry provides. This allows, for example, spatially resolved studies of galaxies, where variations in physical properties can be investigated across different morphological components.

In this paper, we present \textsc{lightstack}, a Python package for constructing homogenized photometric data cubes from multi-band imaging. The package extracts matched cutouts, reprojects them onto a common reference frame, stacks them into three-dimensional $(x,y,\mathrm{filter})$ data products, and performs point spread function (PSF) matching across filters. Although optimized for JWST, HST, LSST, and J-PAS imaging, the workflow of \textsc{lightstack} is survey-agnostic and can be applied to any multi-band FITS dataset with World Coordinate System (WCS) information and PSF specifications.

\section{Pipeline overview}

\begin{figure*}[ht!]
\plotone{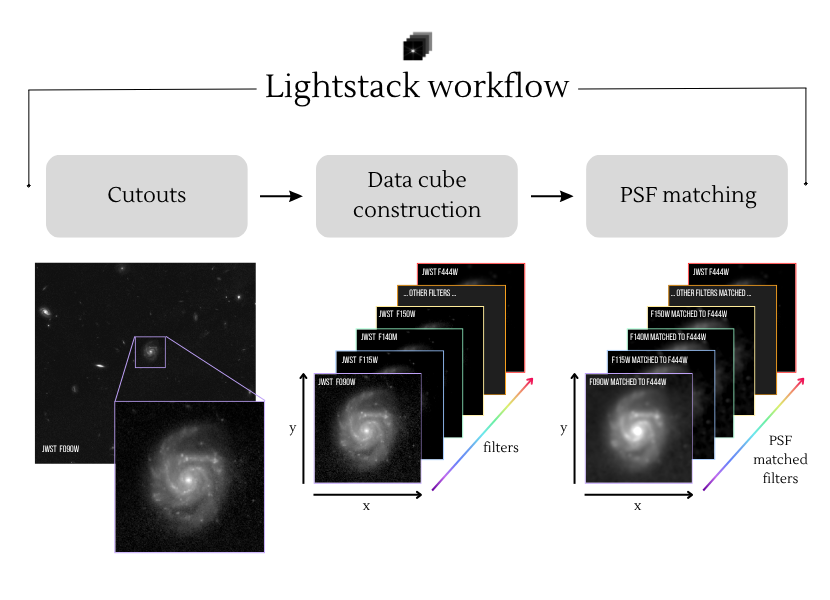}
\caption{Workflow of the code. The example illustrates the steps for constructing a PSF matched photometric data cube, in this case using JWST images. In the first step, a region of interest is cropped: a spiral galaxy observed in the F090W filter. This cropping must be repeated for all available filters. In the second step, all cutouts are stacked in order of wavelengths, resulting in a data cube with dimensions ($x$, $y$, filters). And in the third step, the final output is completed by homogenizing the PSF across all filters. 
\label{fig:workflow}}
\end{figure*}

\textsc{lightstack} was designed to build photometric data cubes in three steps: cropping a region of interest across all available filters; stacking the images to construct the data cube; and performing PSF matching, as shown in Figure \ref{fig:workflow}. In addition to the main functions, the package also provides auxiliary tools that can be used throughout different stages of the analysis, including visualization functions.

\subsection{Cutouts}

Regions of interest, such as individual galaxies, can be extracted from the original mosaics in three different ways. The first method relies on a box region specified in a \texttt{.reg} file.
This approach is appropriate when the mosaic is visually inspected using software such as \textsc{SAOImageDS9}, allowing the user to manually define a box-shaped region enclosing the target. 
The region must be saved to a \texttt{.reg} file and used as input in \textsc{lightstack}. It is a better approach for a more customizable cutout. The second method crops a square region of user-defined size centered on the target coordinates, making it suitable for large samples with cataloged object positions. The last procedure to make cutouts is cropping based on an existing FITS file. This method can be used when a region of interest has already been cropped out in one filter and the same region is required across additional bands, which is useful when working with externally generated cutouts. The algorithm uses the WCS and footprint of this reference FITS file to reproduce an identical cutout in other filters.

\subsection{Data cube construction}

To build a photometric data cube from the extracted regions, the following steps are required. First, all images are aligned and reprojected to a common reference frame. This step requires selecting a reference FITS file (i.e., a specific filter, ideally the one with a larger pixel scale), to which all other images will be matched. After alignment, the images are stacked to construct a 3D data cube with dimensions ($x$, $y$, filter). The correct approach is to use the filters in ascending order of wavelengths. Information from the original 2D FITS header is preserved in the final cube and additional metadata is included (list, order and number of filters, and the reference filter used for alignment). Invalid filters (e.g., empty regions or partially corrupted data) can be removed from the data cube.

\subsection{PSF matching}

The final step ensures spatial resolution homogeneity across all filters through PSF matching. This step assumes that the empirical PSFs for each filter are already available as FITS files. First, convolution kernels are constructed in Fourier space using Fast Fourier Transforms, by selecting a reference PSF (ideally the broadest one). Each filter in the data cube is then convolved with its corresponding kernel to match the reference PSF.

\section{Conclusions}

As multi-band observations and surveys become more frequent, \textsc{lightstack} provides a framework for constructing homogenized photometric data cubes, being an important tool for preparing data. This package has been used in the preparation of datasets for a study on the segmentation and spatially resolved analysis of galaxies \citep{deSouza2026} using James Webb Space Telescope Advanced Deep Extragalactic Survey \citep[JADES;][]{rieke2023, eisenstein2023a, eisenstein2023b} data.

\textsc{lightstack} (v0.2.1) is available on \href{https://github.com/AndressaWille/lightstack}{GitHub} and on Zenodo
\citep[\href{https://doi.org/10.5281/zenodo.20360028}{DOI: 10.5281/zenodo.20360028};][]{lightstack}.

\begin{acknowledgments}

AW acknowledges support from the Coordenação de Aperfeiçoamento de Pessoal de Nível Superior (CAPES) – Funding Code 001.
ACS acknowledges support from FAPERGS (grants 23/2551-0001832-2 and 24/2551-0001548-5), CNPq (grants 312940/2025-4, 445231/2024-6, and 404233/2024-4), and CAPES (grant 88887.004427/2024-00). 

\end{acknowledgments}

\facilities{HST, JWST, LSST, J-PAS.}

\software{astropy
\citep{2013A&A...558A..33A,2018AJ....156..123A,2022ApJ...935..167A},  
          reproject \citep{2020ascl.soft11023R},
          matplotlib \citep{2007CSE.....9...90H}.
          }

\bibliography{sample701}{}
\bibliographystyle{aasjournalv7}

\end{document}